# Effects of Iodine Annealing on Fe$_{1+y}$Te$_{0.6}$Se$_{0.4}$

Jingting Chen, Yue Sun, Tatsuhiro Yamada, Sunseng Pyon, and Tsuyoshi Tamegai

*Department of Applied Physics, The University of Tokyo, 7-3-1 Hongo, Bunkyo-ku, Tokyo*

*113-8656, Japan*

## Abstract

Effects of iodine annealing to induce bulk superconductivity in Fe$_{1+y}$Te$_{0.6}$Se$_{0.4}$ have been systematically studied by changing the molar ratio of iodine to the sample and annealing temperature. The optimal condition to induce bulk superconductivity with $T_c$ ~14.5 K and self-field $J_c$(2 K) ~ 5×10$^5$ A/cm$^2$ is found to be a molar ratio of iodine of 5-7 % at the annealing temperature of 400 °C. Furthermore, the fact that no compounds containing iodine are detected in the crystal and a significant amount of FeTe$_2$ is produced after the iodine annealing strongly indicate that the excess iron is consumed to form FeTe$_2$ and iodine works as a catalyst in this process.

## 1. Introduction

The discovery of the iron-based superconductor (IBS) LaFeAs(O,F) with the superconducting transition temperature $T_c$ ~26 K has stimulated considerable interest in superconductors with layered structures containing iron [1]. Nowadays, IBS families are usually divided into four main types according to their crystal structure: 1111 type [LaFeAs(O,F)] [1], 122 type [(Ba,K)Fe$_2$As$_2$] [2], 111 type [LiFeAs] [3], and 11 type [Fe(Te,Se)] [4]. Among the IBSs, Fe$_{1+y}$Te$_{1-x}$Se$_x$ (11 type) has the simplest crystal structure, consisting of only iron chalcogenide layers, which is advantageous for probing the mechanism of superconductivity [5]. In addition, the 11 type is less toxic than iron pnictides and has small anisotropy [6], which is favorable for applications with high current-carrying capabilities. However, the excess iron that is located at the interstitial site in the Te/Se layer [7] has strong magnetism and acts as pair a breaker and suppresses superconductivity [8,9]. Thus, the as-grown Fe$_{1+y}$Te$_{1-x}$Se$_x$ single crystals only show filamentary superconductivity [10,11].

To induce bulk superconductivity, it is crucial to remove the excess iron. The successful removal of the excess iron from the second Fe site [12] by several different methods has been reported. Until now, the optimal conditions of annealing in atmospheres of O$_2$ [13], S, Se [14], Te [12,15], P, As, and Sb [16,17] have been reported. However, some elements were only reported to be effective in annealing without carefully studying the optimal condition, such as I (iodine) [13,18] and F [19]. On the other hand, iodine has a low boiling point (~184.3 °C), which is advantageous to supply the gas atmosphere for annealing. Furthermore, since iodine is heavy and not contained in the crystal, it is easily detectable on the surface of the crystal. In this paper, the optimal condition for iodine annealing and the mechanism of iodine annealing to induce bulk superconductivity in $Fe_{1+y}Te_{0.6}Se_{0.4}$ are reported.

## *2.* Experiments

$Fe_{1+y}Te_{0.6}Se_{0.4}$ single crystals were grown in the same manner as described in Ref.13. After the growth, the crystals were cut into rectangular shapes with thicknesses of 20~30 μm. For the iodine annealing, crystals and iodine chips (99.9%) were placed together and then sealed in an evacuated quartz tube. However, owing to the temperature dependence of the saturated vapor pressure of iodine, it was problematic that iodine would be vaporized during the evacuation if the temperature was higher than 260 K, causing a problem in confirming the actual amount of iodine that reacted with the crystals. Therefore, the bottom part of the quartz tube containing the iodine and crystals was immersed into a mixture of ice and salt to cool it to 258 K, as shown in Fig. 1(a). After sealing the quartz tube under vacuum condition, it was placed in a furnace at annealing temperatures between 200 and 500 °C for a fixed annealing time (168 h), which was followed by quenching in cold water.

To measure $T_c$ and the critical current density, $J_c$, of the annealed crystal, magnetization measurements were performed using a commercial superconducting quantum interference device (SQUID) magnetometer (MPMS-XL5, Quantum Design). By-products on the sample surface were identified by X-ray diffraction (XRD). Differences in the surface color of the crystal before and after the annealing were inspected using a polarizing microscope. The compositions of the surface of the annealed crystal and the unknown orange material that attached to the inner wall of the quartz tube after annealing were determined by energy-dispersive X-ray spectroscopy (EDX).

## 3. Results and Discussion

Figures 1(b) and 1(c) show a comparison of the sample surface before and after iodine annealing, respectively. The as-grown crystal shows a mirrorlike surface, while some impurities are attached to the surface of the crystal after the iodine annealing at 400 °C. In order to obtain insight into the effect of iodine annealing, we performed XRD measurements of the crystals annealed at 400 °C with different molar ratios of iodine to the crystal. As shown in Fig. 2, the XRD pattern shows only (00$l$) peaks for the as-grown crystal, while some small impurity peaks other than those of $Fe_{1+y}Te_{0.6}Se_{0.4}$ appear, which are assigned to $FeTe_2$ and $FeSe_{1-x}Te_x$ ($x\sim0$).

Figure 3(a) shows the temperature dependences of zero-field-cooled (ZFC) and field-cooled (FC) magnetization for $Fe_{1+y}Te_{0.6}Se_{0.4}$ annealed at 400 °C. The numbers in the legend represent the molar ratio of iodine to the crystal. Here, we define $T_c$ as the onset of diamagnetism. Figure 3(b) shows the magnetic field dependence of $J_c$ for $Fe_{1+y}Te_{0.6}Se_{0.4}$, evaluated from the magnetic hysteresis loops using the extended Bean model [20]. Figure 3(c) summarizes the evolution of $T_c$ and the self-field $J_c$ (2 K) with the molar ratio of iodine. $T_c$ reaches the highest value of 14.5 K when annealing in 6.7% iodine and $J_c$ is up to $5\times10^5$ $A/cm^2$ after annealing in 5.6% iodine under a self-field at 2 K, which is larger than those reported for Fe(Te,Se) [10, 21-26] and similar to the highest reported value for oxygen-annealed $Fe_{1+y}Te_{0.6}Se_{0.4}$ [13, 27]. It has been proved that such a large $J_c$ cannot be obtained from the superconductivity only near the surface [13], which indicates that the sample maintaining such a large $J_c$ obtained in our experiment must be a bulk superconductor. On the other hand, bulk superconductivity cannot be induced by annealing at 400 °C with a molar ratio of iodine of less than 2% (data not shown), and a 4%-iodine-annealed crystal shows small values of $T_c$ and $J_c$ [Fig. 3(c)], different from the oxygen annealing effect [28], where only 3% oxygen annealing can induce bulk superconductivity. Moreover, with increasing amount of iodine, crystals with the highest $T_c$ and largest $J_c$ are obtained when the molar ratio of iodine is 5-7%. Beyond this value, $T_c$ and $J_c$ are suppressed by adding more iodine. The following scenario may explain this behavior. A small amount of iodine can effectively remove excess iron, whereas with increasing amount of iodine, not only the excess iron but also the iron in the Fe(Te,Se) layer will react gradually, causing the decrease in $T_c$ and $J_c$. This scenario is also supported by the fact that when the crystals are annealed with a large amount of iodine (100% iodine) at 400 °C, the surface of the annealed crystal is completely broken and the crystal shows no superconductivity (data not shown).

Next, we continue to study the effect of temperature on annealing. With the decrease in the annealing temperature, peaks from $FeTe_2$ and $FeSe_{1-x}Te_x$ ($x\sim0$) are still detected in the XRD patterns of crystals annealed in a wide range of molar ratios of iodine (data not shown). Figures 4(a) and 4(c) show the results of $T_c$ and $J_c$, respectively, for the crystals annealed at 200 °C, Similarly, Figs. 4(b) and 4(d) show corresponding results for crystals annealed at 300 °C. Although the crystals annealed at both 200 and 300 °C show superconductivity, $T_c$ and the transition width of the crystals annealed at 200 °C [Fig. 4(a)] are lower and broader, respectively, than those of the crystals annealed at 300 °C [Fig. 4(c)], and the difference in $J_c$ for these crystals annealed at 200 and 300 °C is very large [Figs. 4(b), and 4(d)]. Thus, we can conclude that annealing crystals at 300 °C is more effective in removing excess iron than annealing at 200 °C. Figures 5(a) and 5(b) summarize the evolution of $T_c$ and the self-field $J_c$ (2 K) as functions of the molar ratio of iodine to crystal at three different annealing temperatures. Note that the crystals annealed at 200 and 300 °C show $T_c$ lower than 13 K, whereas the crystals annealed at 400 °C usually manifest a higher $T_c$ and a larger $J_c$.

We have presented the results of iodine annealing at temperatures not exceeding 400 °C. In order to determine the optimal annealing temperature, it is worth confirming the superconducting properties of the crystal annealed at temperatures above 400 °C. The result is shown in Fig. 6. Although the crystal annealed at 450 °C with 5.6% iodine shows superconductivity, the superconducting properties are inferior to those of the crystal annealed at 400 °C [Fig. 6(a)]. A similar phenomenon is also observed when the annealing temperature is increased to 500 °C with 20% iodine [Fig. 6(b)].

To probe the mechanism of iodine annealing, identifying the by-products is crucial. However, the information from single crystals is insufficient to fully confirm the by-products. Therefore, we chose two crystals that are annealed under different conditions near the optimal molar ratio and performed XRD after crushing the annealed crystals into powders. Compared with the single-crystal XRD data, more peaks are detected in the powder XRD results shown in Fig. 7. Here, peaks from $FeTe_2$ can be detected not only in the crystals annealed at 200 °C (6.7% iodine) but also in those annealed at 400 °C (10% iodine), which confirmed that $FeTe_2$ is the main by-product after the iodine annealing. The tellurium in $Fe_{1+y}Te_{0.6}Se_{0.4}$ reacts with excess iron and thus leaves mostly selenium and iron, which form $FeSe_{1-x}Te_x$ ($x$~0). Considering the fact that both $FeSe_{1-x}Te_x$ ($x$~0) and $Fe_{1+y}Te_{0.6}Se_{0.4}$ have the same crystal structure, it makes sense that $FeSe_{1-x}Te_x$ ($x$~0) is found only in the XRD data of the $Fe_{1+y}Te_{0.6}Se_{0.4}$ single crystal (Fig. 2), while it disappears in the powder XRD data (Fig. 7). We should also note that no compounds containing iodine are detected in the crystal. Thus, iodine may have served as a catalyst and removed excess iron by forming $FeTe_2$ or other materials rather than reacting directly with excess iron. This assumption is also supported by the fact that neither iodine nor iodine-containing compounds are found in the annealed crystals by EDX, and an unknown orange material that attached to the inner wall of the quartz tubes after annealing was discovered in most of the crystals annealed at different molar ratios of iodine. It has been proved that the unknown material contains iron, tellurium, and iodine by carrying out EDX analysis (data not shown). In this situation, iodine acts as a catalyst as well as a transporting agent [29-31] that may have carried a certain amount of $FeTe_2$ to form the unknown material on the wall of the quartz tube.

## 4. Summary

We have studied the effect of iodine annealing on $Fe_{1+y}Te_{0.6}Se_{0.4}$ single crystals. The iodine-annealed $Fe_{1+y}Te_{0.6}Se_{0.4}$ crystal attains the highest $T_c$ (14.5 K) and the largest $J_c$ ($5 \times 10^5$ A/cm$^2$) when the molar ratio is 5-7 % under the annealing temperature of 400 °C. Iodine serves as catalyst to form $FeTe_2$ and $FeSe_{1-x}Te_x$ ($x \sim 0$), which are by-products after iodine annealing, to remove excess iron.

**Acknowledgment**

This work was partially supported by a bilateral project between Japan and China supported by Japan Society for the Promotion of Science.

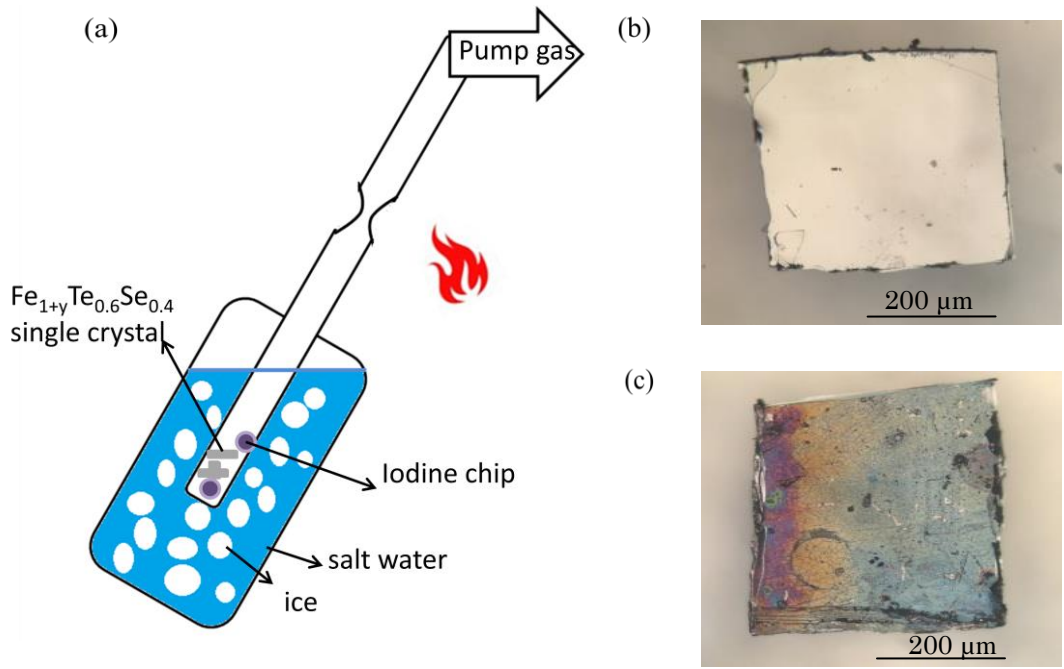

FIG. 1. (color online) (a) Schematic drawing of sealing iodine and $Fe_{1+y}Te_{0.6}Se_{0.4}$ single crystals into a quartz tube immersed in salt water with ice. Optical micrographs of (b) as-grown and (c) iodine-annealed (molar ratio 6.7% and 400 °C) single crystals.

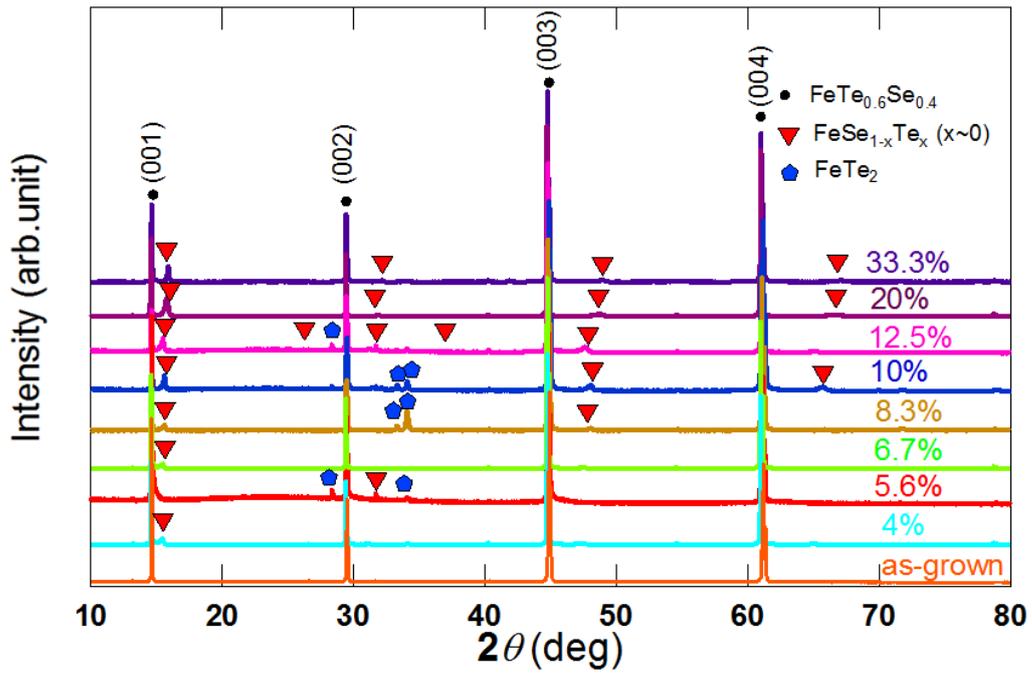

FIG. 2. (color online) XRD patterns of $Fe_{1+y}Te_{0.6}Se_{0.4}$ single crystals annealed at 400 °C with different molar ratios of iodine to the crystal.

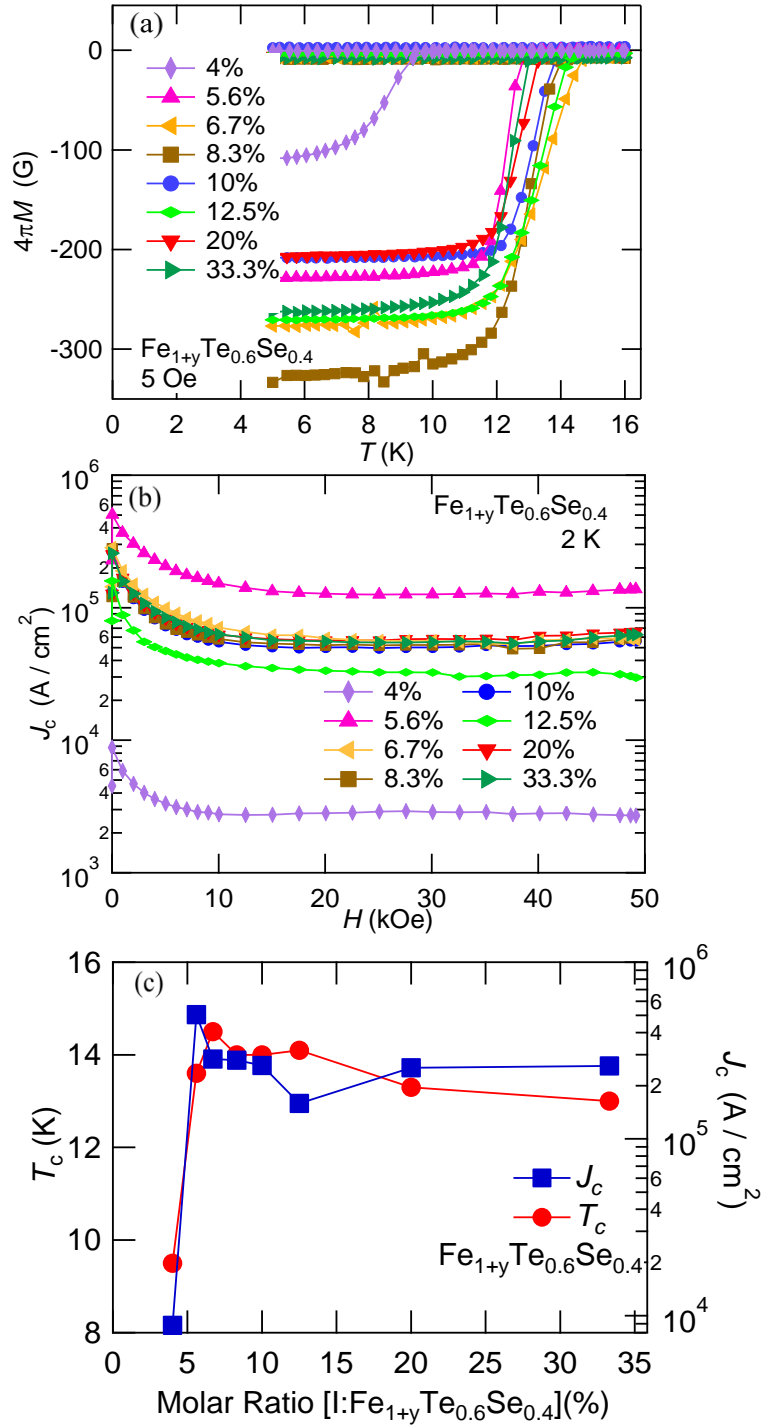

FIG. 3. (color online) (a) Temperature dependences of zero-field-cooled (ZFC) and field-cooled (FC) magnetization and (b) magnetic field dependence of $J_c$ for $Fe_{1+y}Te_{0.6}Se_{0.4}$ annealed with different molar ratios of iodine at 400 °C. (c) Evolution of self-field $J_c$ and $T_c$ for $Fe_{1+y}Te_{0.6}Se_{0.4}$ annealed at 400 °C as a function of the molar ratio of iodine.

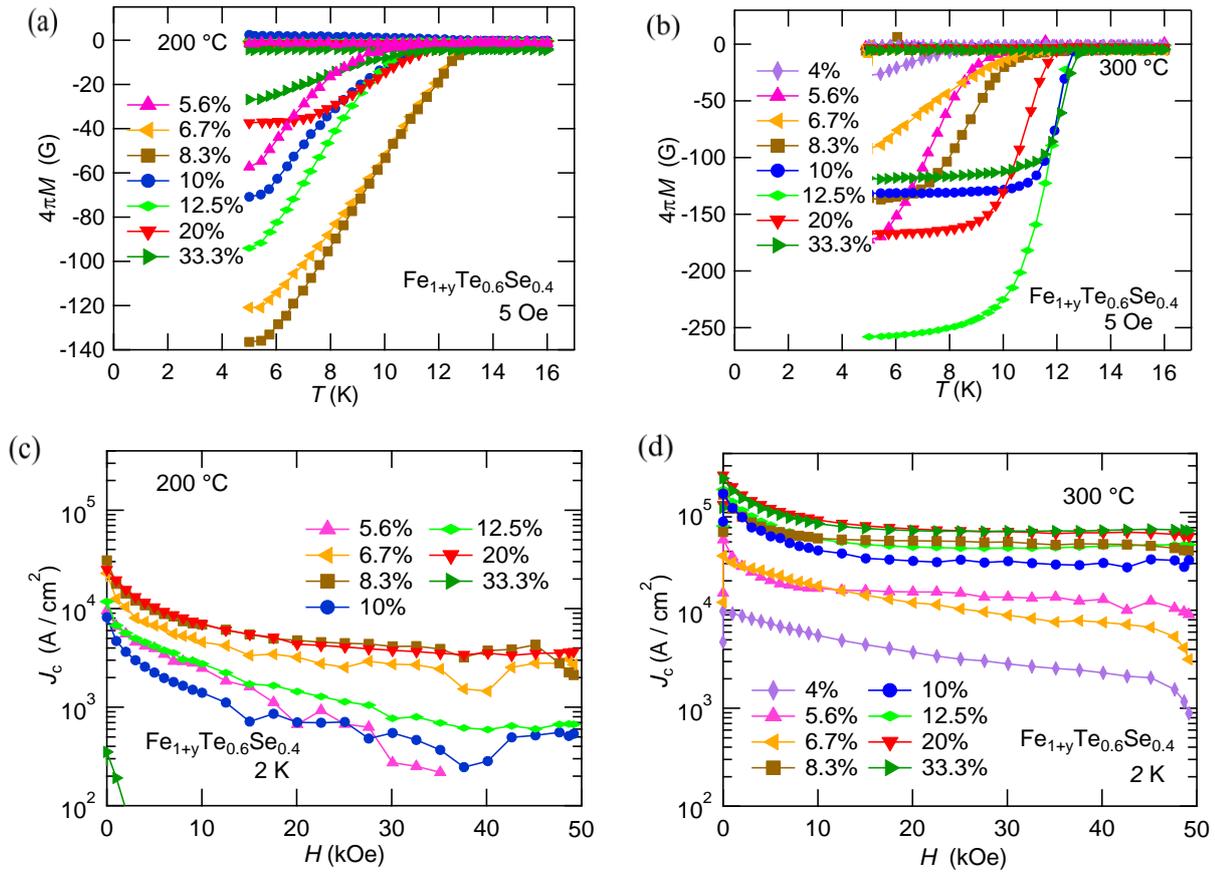

FIG. 4. (color online) (a), (b) Temperature dependence of magnetization and (c), (d) magnetic field dependence of $J_c$ for $Fe_{1+y}Te_{0.6}Se_{0.4}$ annealed at 200 and 300 °C, respectively.

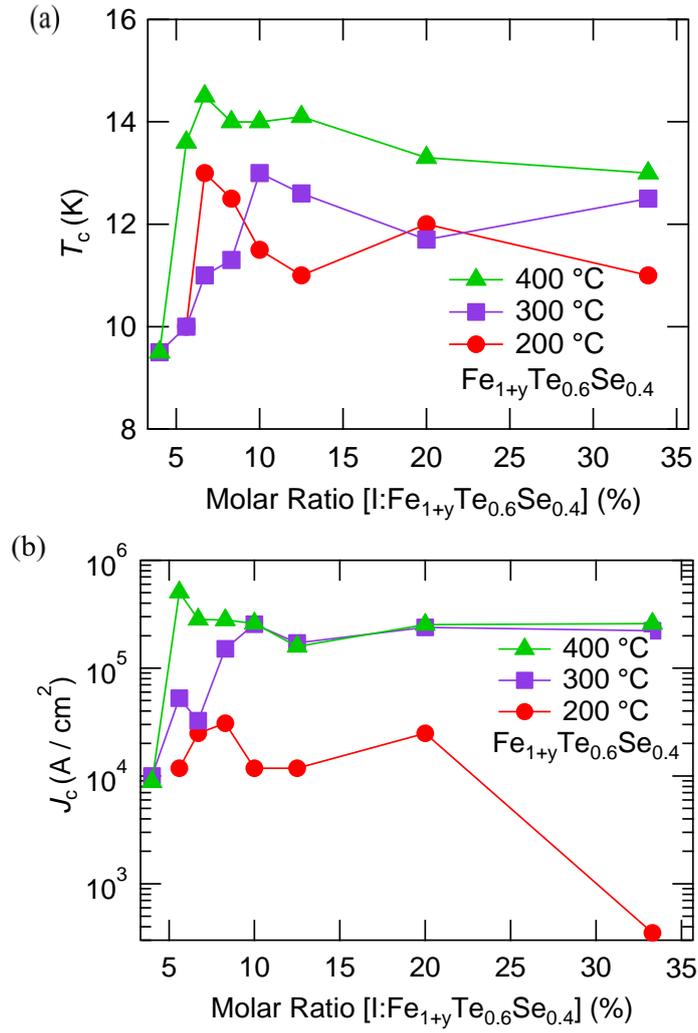

FIG. 5. (color online) (a) $T_c$ and (b) $J_c$ as functions of the molar ratio of iodine for $Fe_{1+y}Te_{0.6}Se_{0.4}$ annealed at different temperatures.

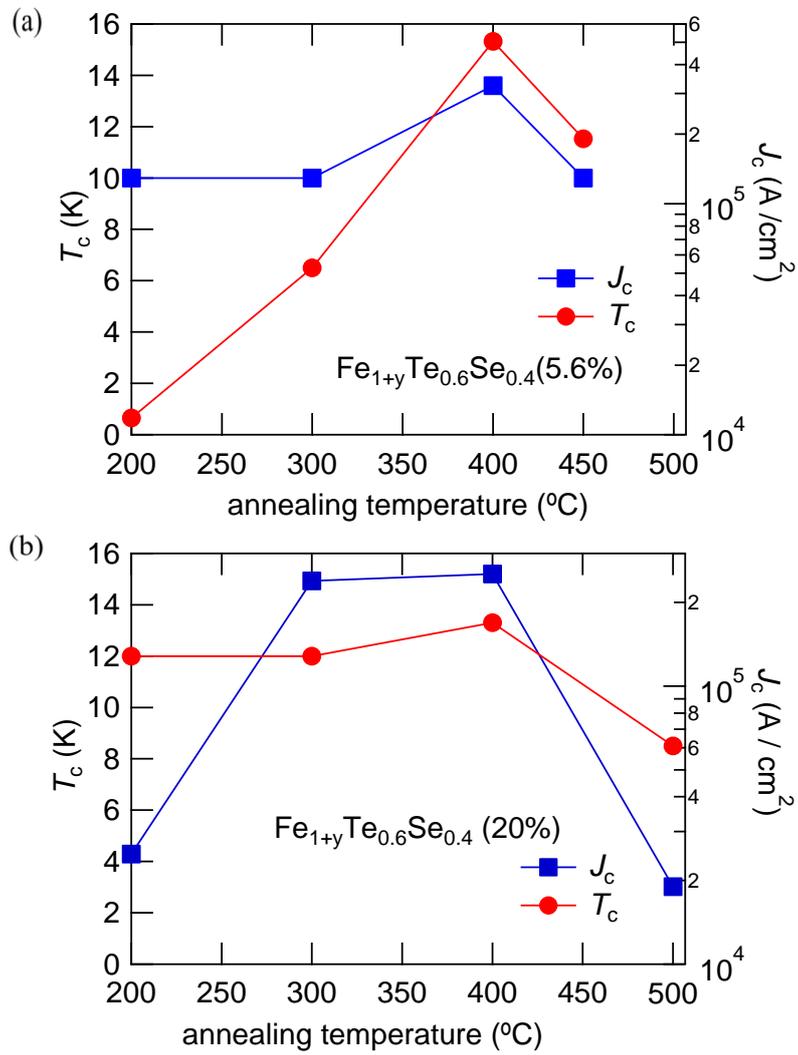

FIG. 6. (color online) Evolution of $T_c$ and $J_c$ as functions of annealing temperature for $Fe_{1+y}Te_{0.6}Se_{0.4}$ annealed with the ratio of iodine to the crystal (a) 5.6 and (b) 20%.

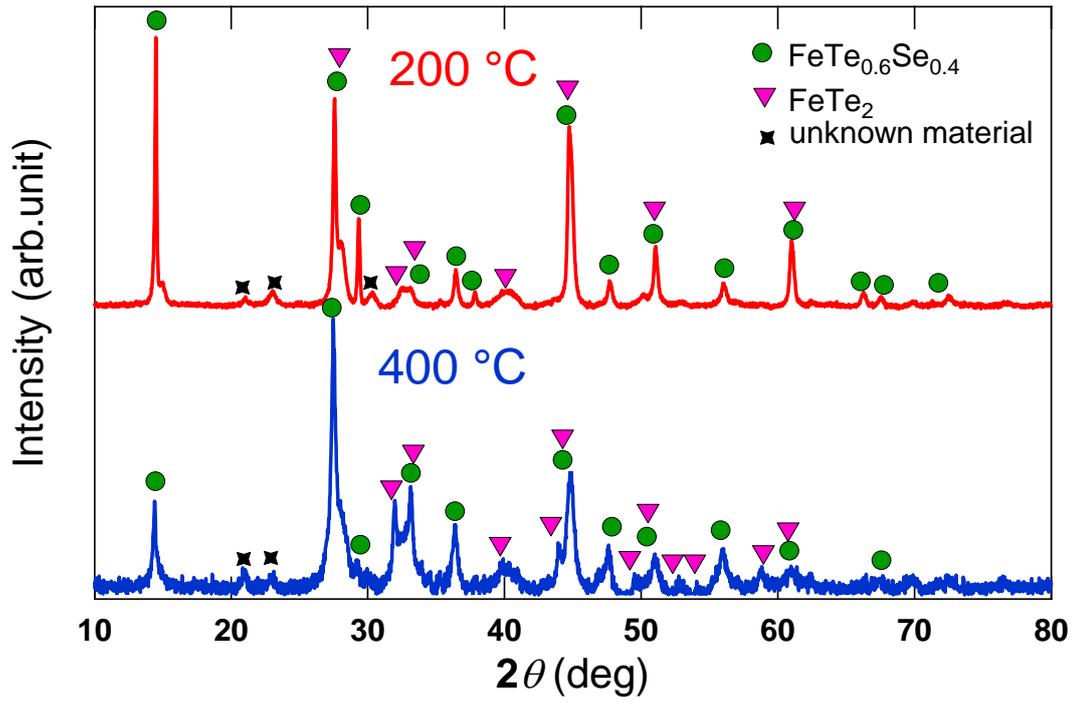

FIG. 7. (color online) Powder XRD patterns for $Fe_{1+y}Te_{0.6}Se_{0.4}$ annealed with 6.7% iodine at 200 °C and 10% iodine at 400 °C.